\documentclass[reprint,aps,prl,superscriptaddress,english]{revtex4-1}
\usepackage{graphicx}
\usepackage{SIunits}
\usepackage{amsmath}
\usepackage{hyperref}

\begin{document}

\title{Critical bursts in filtration}

\author{Filippo Bianchi}
\email{fbianchi@ifb.baug.ethz.ch}
\affiliation{Computational Physics for Engineering Materials, Institute for Building Materials, ETH Z\"{u}rich, 8093 Zurich, Switzerland}
\author{Marcel Thielmann}
\email{Marcel.Thielmann@uni-bayreuth.de}
\affiliation{Bayerisches Geoinstitut, University of Bayreuth, 95440 Bayreuth, Germany}
\author{Lucilla de Arcangelis}
\email{lucilla.dearcangelis@unicampania.it}
\affiliation{Department of Industrial and Information Engineering, University of Campania Luigi Vanvitelli, 81031 Aversa (CE), Italy}
\author{Hans J\"{u}rgen Herrmann}
\email{hans@ifb.baug.ethz.ch}
\affiliation{Computational Physics for Engineering Materials, Institute for Building Materials, ETH Z\"{u}rich, 8093 Zurich, Switzerland}

\begin{abstract}

Particle detachment bursts during the flow of suspensions through porous media are a phenomenon that can severely affect the efficiency of deep bed filters. Despite the relevance in several industrial fields, little is known about the statistical properties and the temporal organization of these events. We present experiments of suspensions of deionized water carrying quartz particles pushed with a peristaltic pump through a filter of glass beads measuring simultaneously pressure drop, flux and suspension solid fraction. We find that the burst size distribution scales consistently with a power-law, suggesting that we are in the presence of a novel experimental realization of a self-organized critical system. Temporal correlations are present in the time series, alike in other phenomena as earthquakes or neuronal activity bursts, and also an analog to Omori's law can be shown. The understanding of bursts statistics could provide novel insights in different fields, e.g. in filter and petroleum industries.

\end{abstract}

\maketitle

Filtration of fluids through deep bed filters is a problem of great technological interest due to the wide range of related applications, such as the purification of liquid aluminium for can construction, the design of self-regenerating filters or the prediction of sand production in oil wells. Experimentally it has been shown that, depending on local flow conditions, bursts appear in fluid flow and pressure drop across the filter \cite{Gruesbeck1982,Einstein1992,Sahimi2000}, which represent an important problem affecting the efficiency and lifetime of the filter itself. Despite the relevant technological impact of bursts, no attention has been given so far to a detailed statistical analysis of their occurrence. Such analysis is of great importance, since information about the range of burst sizes and their temporal organization can help the optimization of filter performance. We therefore experimentally address this problem in order to measure and analyze statistical properties of long time series of bursts.

A number of microscopic mechanisms is responsible for deposition and resuspension of particles leading to bursts. Particle deposition reduces filter porosity and permeability, and it occurs  through particle interception by the filter matrix, inertial impaction of suspended particles against the matrix, Brownian diffusion, and gravity \cite{Herzig1970,Sutherland2008,Tien2012}. Upon contact, small particles (\textless 10\textsuperscript{1}~\micro\metre) stick to the filter matrix surface because of Van der Waals forces \cite{Herzig1970,Sutherland2008,Tien2012}, whereas larger particles are mainly retained by mechanical clogging due to particle straining and bridging in pore throats \cite{Herzig1970,Valdes2006}. Conversely, fluid flow induced drag forces are responsible for particle detachment, as soon as they exceed the adhesive forces between the particle and the filter matrix or deposit \cite{Bai1997,Bergendahl2000}. In order to observe particle resuspension \cite{Bai1997,Mahadevan2012} it is necessary to achieve a certain value of fluid velocity, which is then enhanced if flow surges occur \cite{Han2009,Kim2012}. Detachment is also favored by instabilities caused by suspended particles \cite{Ives1989} that hit and detach previously deposited particles or pore pressure fluctuations due to the passage of other particles closeby \cite{Ghidaglia1996}. Resuspended particles can then either be re-entrapped in deeper filter layers or exit the filter with the effluent \cite{Bai1997,Kim2004}.

Experimental investigations have shown that a dynamic equilibrium can be reached between deposition and detachment rates \cite{Gruesbeck1982}. Such equilibrium corresponds to morphological modifications inside the filter \cite{Imdakm1987} caused by the plugging and unplugging of pores. This effect becomes visible as bursts, i.e. fluctuations either in the solid fraction (ratio between volume of solid particles and total volume) of the effluent \cite{Gruesbeck1982} or in the differential pressure or else in the permeability through the filter \cite{Einstein1992} (which is observed at the field scale in oil reservoirs \cite{Sahimi2000}). Several models have therefore been developed to take into account deposition and detachment of particles (see e.g. Ref. \cite{Imdakm1991,Sahimi1991,Ohen1993,Ju2007,Civan2007,Ochi1999,Sahimi1990}). In this study we characterize the statistical properties of burst sizes by experimentally generating long time series. We also question the existence of temporal correlations between events which could lead to a sequence of close-in-time bursts seriously affecting the filter.

We measure the flow of a Newtonian aqueous suspension of quartz particles (whose size is large enough for Brownian motion to be negligible) through a densely packed filter bed of glass beads (see Fig. \ref{Fig1} and the Supplemental Material for details). The closed loop design of our setup allows us to perform long-term experiments, thus to collect long data series necessary to perform a reliable statistical analysis. In a certain range of experimental parameters, we observe permeability jumps which are related to particle resuspension bursts inside the filter.

\begin{figure}[h!]
\includegraphics[width=\linewidth]{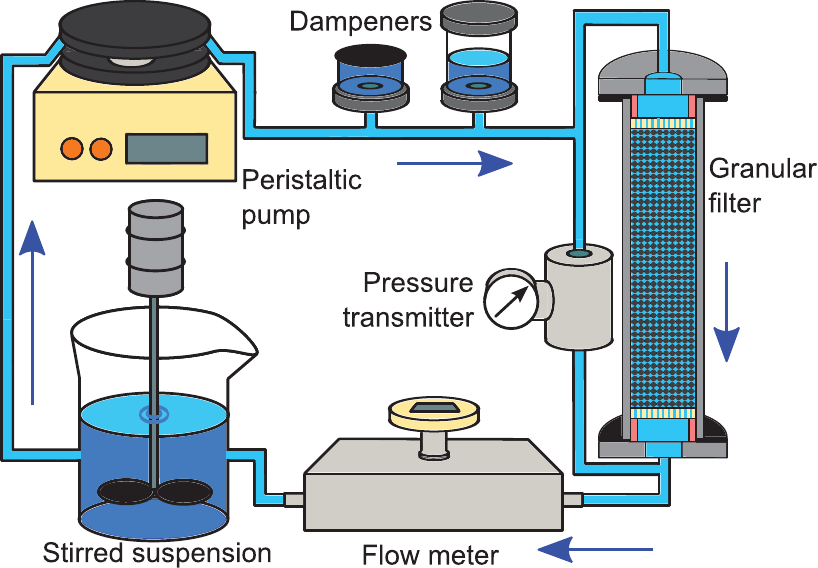}
\caption{Experimental setup. The suspension is pumped from a beaker through two pressure oscillation dampeners and through the granular filter.}
\label{Fig1}
\end{figure}

\paragraph{Filtration regimes.}

The clogging behaviour of the filter is strongly influenced by the solid fraction ($\Phi=\left(\rho-\rho_w\right)/\left(\rho_s-\rho_w\right)$, where $\rho$ is the suspension density, $\rho_w$ the water density, and $\rho_s$ the quartz density) whereas the flow rate $Q$ has a negligible effect (see Fig. \ref{Fig2}). Three different regimes are identified: A non-clogging regime, a clogging regime, and an intermediate transient regime.

\begin{figure}[ht]
\includegraphics[width=\linewidth]{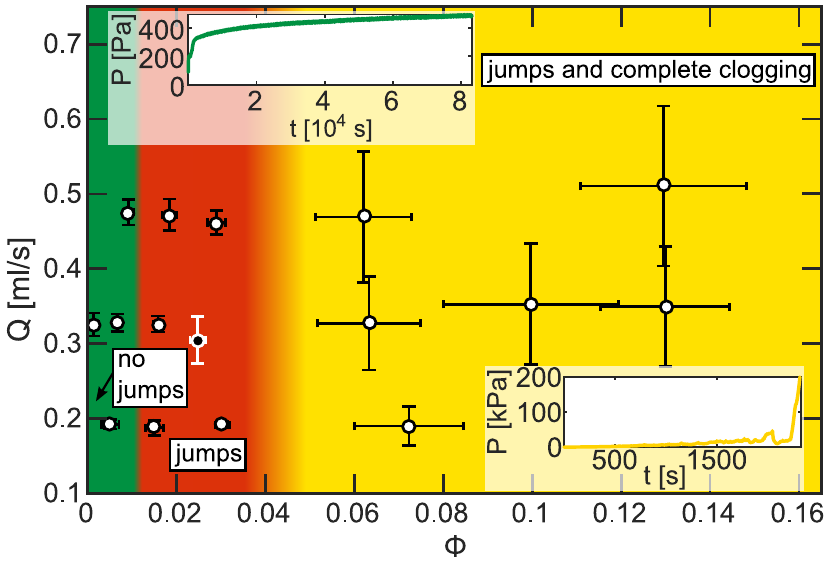}
\caption{Filter phase diagram. At low solid fractions the flux is continuous with no indications for clogging (green area). With increasing solid fraction, jumps are observed (red area). For even higher values of $\Phi$, the filter reaches complete clogging (yellow area). Every point in the diagram indicates an experiment, error bars represent variations of $Q$ and $\Phi$ ($\pm1.96\sigma$, $\sigma$ is the data standard deviation). The black point with white error bars indicates the parameters of the experiments whose statistical analysis is shown in the following. Insets show typical pressure loss vs. time evolution for experiments of non-clogging (upper inset) and clogging regime (lower inset).}
\label{Fig2}
\end{figure}

At low solid fractions (\textless ~10\textsuperscript{-2}) pressure loss through the filter ($P=P_1-P_2$, where $P_1$ and $P_2$ are the pressures at the filter inlet and outlet respectively) rises very slowly, approaching asymptotically a constant (upper inset of Fig. \ref{Fig2}). No significant fluctuations are observed, indicating that deposition does not play an important role and that particle detachment bursts are absent.

\begin{figure}[h!]
\includegraphics[width=\linewidth]{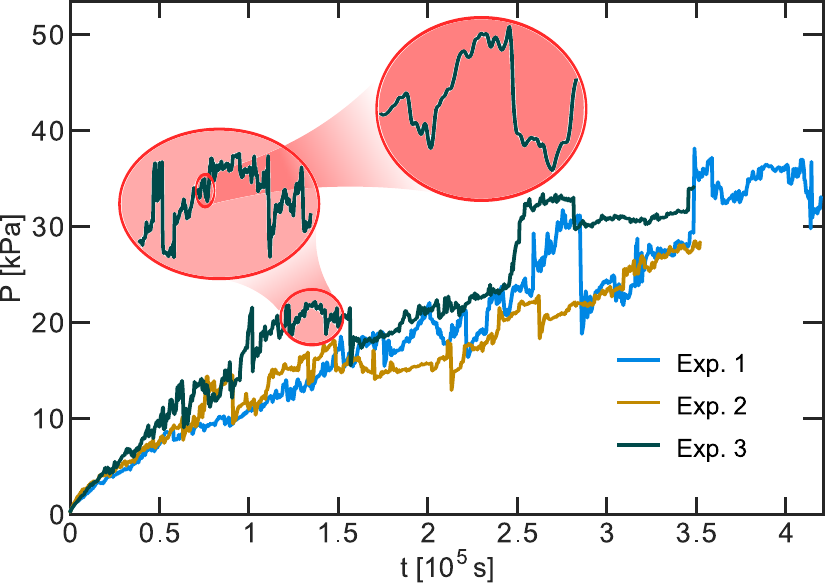}
\caption{Time evolution of pressure loss $P$ through the filter during three experiments. Successive zooms into a temporal interval are shown in the balloons.}
\label{Fig3}
\end{figure}

Experiments performed with a suspension solid fraction larger than 10\textsuperscript{-2} show a series of pressure loss jumps, thus in permeability. If $\Phi$ is smaller than ~4$\cdot$10\textsuperscript{-2}, complete clogging does not occur. The overall rise of pressure loss is very slow (3$\cdot$10\textsuperscript{4}~\pascal\: in more than 4 days for the experiments shown in Fig. \ref{Fig3}). This implies that experiments are close to a steady state, in which the mean value of $P$ is stationary and only fluctuates due to pressure loss jumps. This is an indication of a dynamical competition between deposition and resuspension inside the filter.

If $\Phi$ is increased above $\simeq$~5-6$\cdot$10\textsuperscript{-2}, pressure loss jumps are still observed during the experiments (lower inset of Fig. \ref{Fig2}). In this case, the filter permeability reduction is so strong that complete clogging is reached and fluid flow through the filter stops completely.

\paragraph{Resuspension bursts.}

\begin{figure}[h!]
\includegraphics[width=\linewidth]{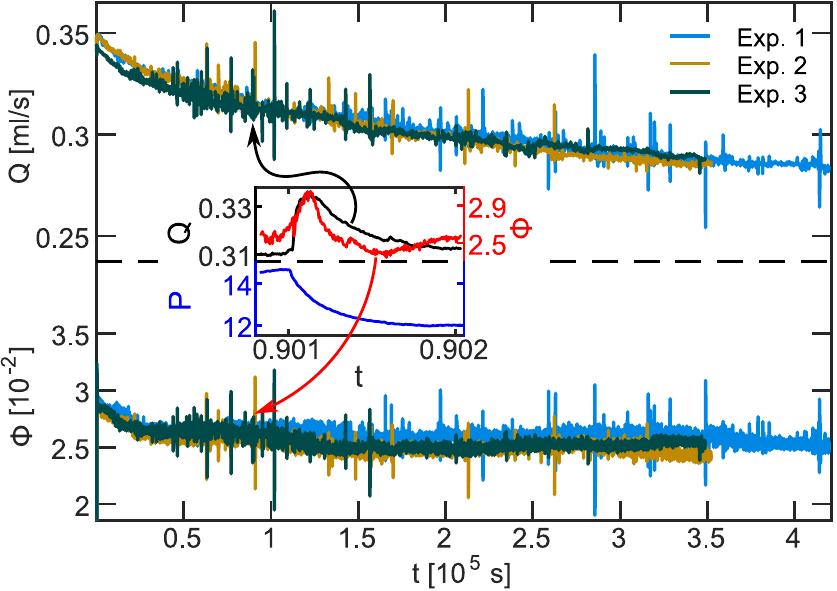}
\caption{Time evolution of the flow rate $Q$ and the suspension solid fraction $\Phi$ during the three experiments. The inset shows the temporal evolution of $Q$ and $\Phi$ (top), and $P$ (bottom) during a single jump of exp. 3 indicated by the arrows (units are the same as in the main figure, $P$ is in \kilo\pascal).}
\label{Fig4}
\end{figure}

Here we analyze three experiments run under the same experimental conditions (Fig. \ref{Fig2}) in the transient regime for about 4.5~days. The temporal evolution of the pressure loss, the flow rate and the suspension solid fraction are measured as a function of time (Figs. \ref{Fig3} and \ref{Fig4}).

Figure \ref{Fig3} shows that $P$ increases in time with a sequence of successive jumps which can be attributed to particle deposition inside the filter. The time evolution of the flow rate also exhibits jumps, which occur at the same time as the pressure loss jumps (Fig. \ref{Fig4}). Indeed the flow rate increases at the beginning of a jump, reaches a maximum and then decreases to its initial value (inset of Fig. \ref{Fig4}). At the same time the suspension solid fraction follows the same behaviour as the flow rate, with a faster decay to the initial value.

The fact that $Q$ increases together with pressure loss jumps implies that sample permeability increases as well, which can be explained by the opening of new channels due to resuspension of previously deposited particles. The frequent occurrence of these jumps indicates that experiments are characterized by a continuous interplay between deposition phases and detachment bursts. When $P$ increases, deposition dominates and pores in the filter become increasingly clogged. Consequently, local flow rates in the individual pores increase. As soon as a sufficient flow rate is reached, particle detachment occurs, resulting in the opening of a pore. Permeability then increases, while $P$ and the local fluid velocity decrease, thereby restarting the cycle.

We compute the jump abruptness as the ratio between pressure loss jump size $\Delta P$ and its duration $\tau$ (see Supplemental Material for details on jump recognition). A log-log  scatter plot of the jump abruptness $\Delta P/\tau$ vs. $\Delta P$ for all data shows a linear trend (Fig. S2 on Supplemental Material). Despite the scatter in the data, results are consistent for all three experiments and indicate that large jumps experience a faster decrease in pressure loss than smaller ones. A linear fit to the data suggests that jump abruptness scales as the jump size to the power of 0.73 $\pm$0.03. A discussion about the mechanism behind this phenomenon can be found in the Supplemental Material.

\paragraph{Size and duration distributions of jumps.}

\begin{figure}[h!]
\includegraphics[width=\linewidth]{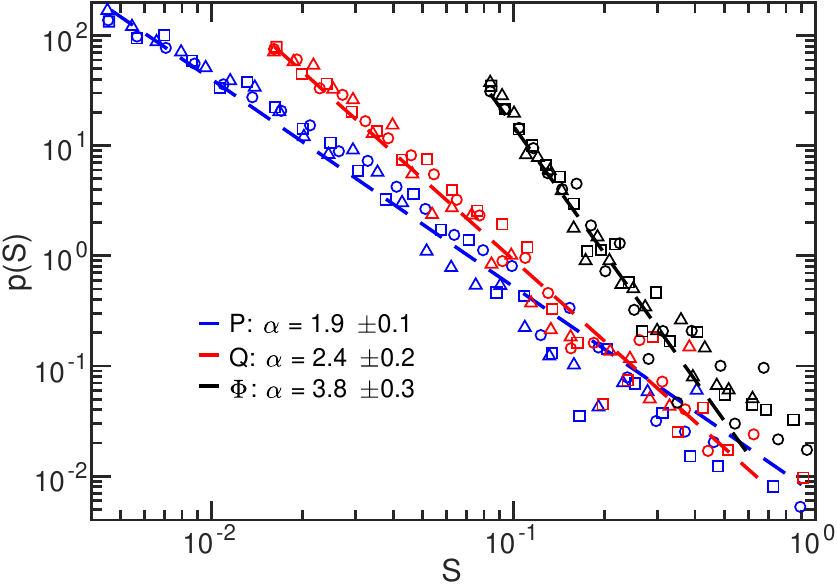}
\caption{Size distribution of pressure loss (blue), flow rate (red), and solid fraction (black) jumps. Circles are data from exp. 1, triangles from exp. 2, squares from exp. 3. The jump size $S$ is normalized by the size of the largest jump. The error of the exponents $\alpha$ is evaluated as the maximum difference between the fitted value for data from all experiments and from single experiments.\label{Fig5}}
\end{figure}

Jump sizes vary significantly during a single experiment (Fig. \ref{Fig5} and Supplemental Material for details), and their distributions are consistent with power-laws with exponents $\alpha$~=~1.9~$\pm$0.1 for $P$, 2.4~$\pm$0.2 for $Q$ and 3.8~$\pm$0.3 for $\Phi$. The exponents are computed applying the maximum likelihood method \cite{Newman2005} to data from three experiments. The power-law behaviour of pressure loss jump size distribution is confirmed by numerical simulations of Ref. \cite{Jaeger2017}. Moreover we measure the pressure loss jump size distribution by applying a smaller identification threshold (5 instead of 50~\pascal , see Supplemental Material for details). The power-law behaviour is always retrieved with very stable exponents, implying that the absence of a characteristic size does not depend on the identification threshold. A deviation from the power-law distribution, due to the loss of events caused by the limited resolution of acquisition devices is only found for small jump sizes. Because of experimental restrictions (as the necessity of being in the transient regime) that do not allow a strong variation of flowing parameters, the extension of size distributions of $Q$ and $\Phi$ is limited. Interestingly, the values of $\alpha$ for all three distributions are larger than the values measured for earthquakes, solar flares and bursts in neuronal activity, which are typically in the range of 1.5-1.6 \cite{DeArcangelis2016,Mendoza2014,Beggs2003}, implying that filtration belongs to a different SOC universality class.

\paragraph{Temporal correlations.}

\begin{figure}[h!]
\includegraphics[width=\linewidth]{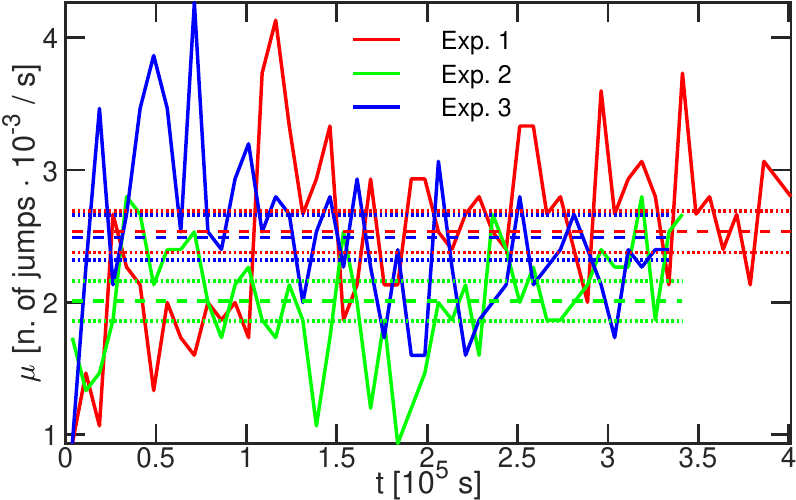}
\caption{Pressure loss jump rates for the three experiments. Dashed lines are the average Poissonian rates, dotted lines are the 95~\% confidence intervals.\label{Fig6}}
\end{figure}

The consistency of size distributions with power-laws suggests that, in the transient regime, filtration exhibits a scale-free behaviour and that the phenomenon is critical. Beside the absence of a characteristic size, a fundamental feature of criticality is the existence of long-range temporal correlations between bursts. In order to verify their existence, we calculate the pressure loss jump rate $\mu$ (over non-overlapping windows of 5000~\second ) as a function of time for each experiment (Fig. \ref{Fig6}). We observe that $\mu$ is not a constant, as one would expect for a perfect Poisson process. The jump rate highly fluctuates, exceeding frequently the 95~\% confidence intervals with respect to the average Poissonian rate for each experiment. The existence of such jumps in the rate suggests the presence of correlations in the time series.

\begin{figure}[h!]
\includegraphics[width=\linewidth]{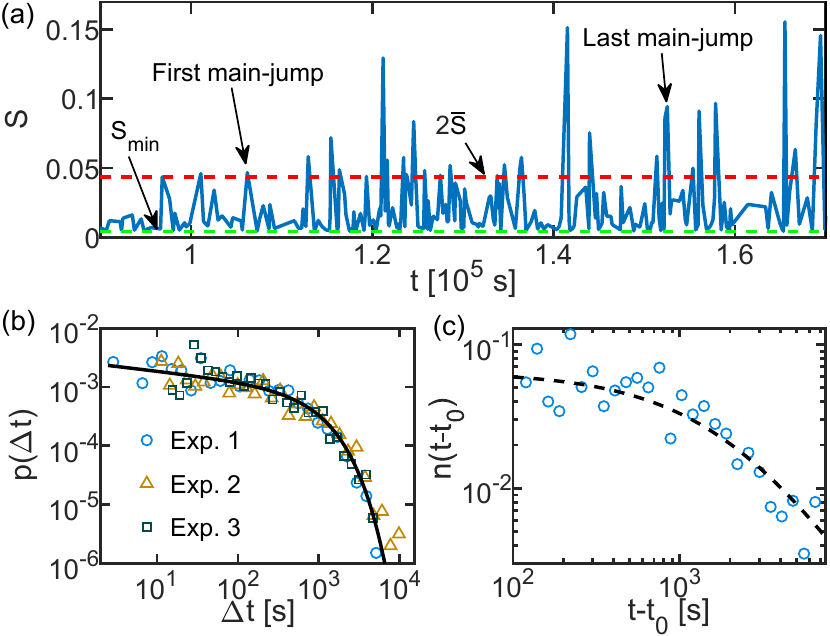}
\caption{(a) Jump series for exp. 1 during a phase in which $\mu$ is higher than average. Main-jumps are events above the red line. The green line indicates the minimum jump size. (b) Quiet time distributions are fitted by a Gamma distribution. (c) The occurrence rate of after-jumps can be fitted with the modified Omori's law. \label{Fig7}}
\end{figure}

In order to confirm the presence of correlations we evaluate the quiet times distributions. The quiet time is the time lag between the end of a pressure loss jump and the beginning of the next one, considering only jumps  larger than 150~\pascal . These distributions, shown in Fig. \ref{Fig7}(b), exhibit a more complex behaviour than the exponential distribution expected for a Poisson process. Indeed the distributions can be fitted by a Gamma function, $p\left(\Delta t\right)=1/\left(B^q\Gamma\left(q\right)\right)x^{q-1}e^{-x/B}$ with $q$~=~0.84~$\pm$0.04 and $B$~=~1006~$\pm$370~\second\: (the error of the fitted parameters $q$ and $B$ represents the maximum difference between the fitted parameters of data from all experiments and the fitted parameters of data from a single experiment). They are therefore well approximated by a power-law with exponent $\beta=1-q$~=~0.16 in an initial regime until $\Delta t\lesssim$~10\textsuperscript{3}~\second , whereas for longer quiet times an exponential decay sets in. To confirm that a Gamma function represents the best fit to the data, we evaluate the Akaike information criterion for a Gamma and an exponential function. We find that the Gamma function has a lower $AIC$ compared to the simple exponential, leading to a relative likelihood $K=e^{(AIC_{gamma}-AIC_{exp})/2}$~=~1.67$\cdot$10\textsuperscript{-5} in favour of the Gamma function. If a lower identification threshold is implemented, the quiet time distribution exhibits an exponential behaviour (see Supplemental Materials), indicating that correlations exist mainly between large events, whereas small fluctuations in the pressure loss are dominated by uncorrelated random noise. This is a further indication that the filtration process in this regime is critical. The evidence for a Gamma function quiet time distribution is an intriguing result since this functional behaviour, with similar values of $\beta$, is common to several natural stochastic processes, such as earthquakes \cite{Corral2003,Corral2004,DeArcangelis2016}, solar flares \cite{Crosby1998,Wheatland1998}, and acoustic emissions in rock fracture \cite{Davidsen2007}, where close-in-time events are temporally correlated. 

Inspired by the statistical properties of seismic sequences, we investigate in deeper detail the temporal organization of events during an interval characterized by large rate increase. We focus on the temporal sequence of events from experiment 1 between 1.1$\cdot$10\textsuperscript{5} and 1.5$\cdot$10\textsuperscript{5}~\second\: [Fig. \ref{Fig7}(a)]. We define the main-jump as a jump with size $S>2\overline{S}$, where $\overline{S}$ is the average jump size in the analyzed time period. Its after-jumps are the following jumps occurring before the next main-jump. This procedure to identify main- and after-events is analogous to standard statistical analysis performed for earthquakes \cite{DeArcangelis2006,DeArcangelis2016}. Figure \ref{Fig7}(c) shows the number of after-jumps in time, $n(t-t_0)$, occurring after a main-jump occurred at $t=t_0$. We observe that $n(t-t_0)$ behaves accordingly to the modified Omori's law of earthquakes \cite{Utsu1995,Main2000} $n\left(t-t_0\right)=A\left(c+t-t_0\right)^{-p}$ with an exponent $p$~=~1.8, and a $c$ value of 2.2$\cdot$10\textsuperscript{3}~\second\: (the parameters are estimated according to Ref. \cite{Ogata1983}). This indicates that $n(t-t_0)$ has a power-law decay with exponent 1.8 (for earthquakes $p\simeq$~1 \cite{Omori1894}). If data are fitted with a pure power-law, the exponent is 1.4 with a smaller error bar. This result suggests that, as large earthquakes trigger a sequence of aftershocks whose rate decreases in time as a power-law, large jumps trigger sequences of smaller close-in-time jumps.

\paragraph{Conclusions.}

Our experiments show that jumps in deep bed filtration are the expression of a self-organized critical process, occurring in a regime of parameters where the balance between deposition and detachment of particles is realized. The values of the critical exponents for the jump size distributions are different than those typically measured for other stochastic natural processes, suggesting that this phenomenon is in a novel universality class of self-organized critical phenomena. Indeed, the absence of a characteristic event size and power-law distributions, even if with different exponents,  are found in a variety of natural phenomena, as earthquakes \cite{Gutenberg1944,DeArcangelis2016}, solar flares \cite{Lu1991,Mendoza2014}, stock markets \cite{Gopikrishnan2000,Bartolozzi2005}, and neural avalanches \cite{Beggs2003,DeArcangelis2014}. Therefore the evidence for a novel universality class suggests that the microscopic mechanisms controlling the self-organization in filtration are different from all the aforementioned processes. Indeed filtration can attain a self-organized critical state by a dynamical adjustment of the porous medium, in which the decrease or increase of local fluid velocity reflect the resuspension or deposition of particles, respectively. This microscopic interpretation has been recently confirmed by numerical simulations that are able to reproduce the scaling behaviour of the size distributions \cite{Jaeger2017}. Moreover, the existence of temporal correlations between close-in-time events is evidenced by the power-law decay of the occurrence rate, similarly to the Omori's law for earthquakes. These correlations between bursts confirm the relevance of hydrodynamic interactions in the process. Our results can be of interest for a number of problems where resuspension events are observed, as for instance sand production in oil wells or filtration of pollutants in soil where resuspension bursts could release contaminants in the effluent.

\begin{acknowledgments}

The research leading to these results has received funding from the European Research Council, ERC Advanced Grant 319968-FlowCCS. We acknowledge the support of Falk K. Wittel and Claudio Madonna for helping in the design and realization of the experiments.

\end{acknowledgments}

\section*{Appendix A: experimental setup}

The aqueous suspension used in the experiments is composed of deionized boiled water and quartz powder (\textgreater ~230 mesh, Sigma Aldrich). The d\textsubscript{50} (median size in the cumulative grain size distribution)  of quartz particles is 25~\micro\metre . The granular packing constituting the filter consists of soda lime glass beads (Sigmund Lindner) with a diameter of 0.9 to 1~\milli\metre . The granular packing is embedded in an aluminium cylinder (height 80~\milli\metre , diameter 16~\milli\metre). A peristaltic pump (Gilson MINIPULS 3), controlled via computer through an I/O module (Meilhaus Electronic RedLab 1208LS), pumps the fluid through the granular packing. Pressure oscillations generated by the pump are dampened by a combination of a neoprene membrane dampener and an air chamber. The pressure loss through the experimental sample is measured with a differential pressure transmitter (Keller PRD-33~X, $\pm$145~\pascal\: accuracy). A Coriolis flow meter (Endress+Hauser Proline Promass A 100) is placed at the sample outlet to measure volumetric ($\pm$0.4~\micro\litre\per\second\: accuracy) and mass ($\pm$0.6~\milli\gram\per\second\: accuracy) flow rates. It allows for the calculation of the suspension solid fraction ($\Phi$), as the volumetric ($Q$) and mass ($\dot{M}$) flow rates, water density ($\rho_w$~=~998.4 - 999.4~\kilo\gram\per\cubic\metre), and quartz density ($\rho_s$~=~2648~\kilo\gram\per\cubic\metre) are known ($\Phi=\left(\rho-\rho_w\right)/\left(\rho_s-\rho_w\right)$, where $\rho=\dot{M}/Q$ is the suspension density). The flow meter is connected to a PC via a DAQ device (Meilhaus Electronic RedLab 1208FS).

\section*{Appendix B: Sample preparation and experimental procedure}

To prepare the sample for the experiments, a 3D printed polycarbonate sieve (3~\milli\metre\: height) was positioned at the lower side of the aluminum cylinder. A rubber ring (6~\milli\metre\: height), placed below the sieve held it in place. The cylinder was then filled with glass beads (20~$\pm$0.05~\gram), placed on a vibrating table (a loud speaker connected to an oscilloscope) and vibrated at 150~\hertz\: for 30~\second . Lids were fastened to the cylinder after placing a polycarbonate sieve and a rubber ring on the upper end. A seal was created by neoprene rings positioned between the lids and the cylinder. The grains were compressed by the polycarbonate filters and the rubber rings, to ensure packing rigidity.

The sample was initially filled with clean water. Air bubbles were removed by flushing the sample in both directions with water at flow rates of $\simeq$0.7~\milli\litre\per\second , while simultaneously shaking it vigorously. Upon removal of all air bubbles, the quartz powder was added to the water and the experiment started at the desired flow rate. During the whole experiment, the pressure loss through the sample, the volumetric flow rate, and the mass flow rate were recorded at a sampling rate of 2~\hertz . At the end of the experiment, the sample was cleaned with deionized water. We produced many samples but all the experiments reported in the letter were run with the same bead packing.

\section*{Appendix C: Data postprocessing: identification of jumps}

Time series of the pressure loss, flow rate, and suspension solid fraction were filtered with a second order low-pass Butterworth filter with a cutoff frequency of 0.015~\hertz\: to remove experimental noise (instrumentation noise and pump pulsations). Jumps are then detected in the filtered data. The size of those jumps is computed as the difference between a local maximum (or minimum for $Q$ and $\Phi$) and the following local minimum (or maximum for $Q$ and $\Phi$). Jumps with sizes smaller than 50~\pascal\: ($P$), 1~\micro\litre\per\second\: ($Q$), 7$\cdot$10\textsuperscript{-4} ($\Phi$) were neglected as their amplitude is smaller than the noise amplitude.

\section*{Appendix D: Jump abruptness}

Jump abruptness scales with jump size to the power of 0.73 $\pm$0.03 ($\Delta P/\tau=0.039\Delta P^{0.73}$ is the straight line of Fig. \ref{Fig1}). The error of the exponent represents the maximum difference between the fitted exponent of the data from all the experiments and the fitted exponent of data from a single experiment.

\begin{figure}[h!]
\includegraphics[width=\linewidth]{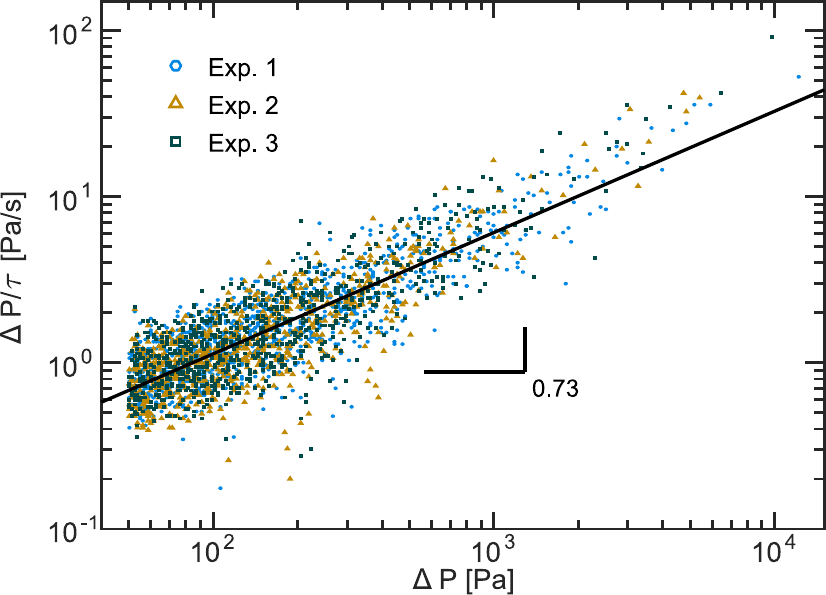}
\caption{Jump abruptness vs. jump size. The black line is the linear fitting on log-log scale, whose slope is 0.73. Every point stems from a pressure loss jump.}
\label{Fig1}
\end{figure}

An explanation of this observation is given in the following. When a wide channel (or more channels at the same time) forms, the filter undergoes a substantial permeability increase. Consequently the pressure loss decreases significantly, resulting in a large jump. Vice versa, if a narrow channel is created the permeability increase will be limited and the related pressure loss jump smaller. A wide channel is also able to discharge fluid at a higher flow rate than a narrow channel, resulting in a faster drop in pressure loss for large jumps.

\section*{Appendix E: Jump size distribution}

Depending on the minimum threshold used to recognize pressure loss jumps, a deviation from the power law distribution may appear for smallest jump sizes (see Fig. \ref{Fig2}). Such jumps would be smaller than the noise due to the experimental apparatus (oscillations coming from the peristaltic pump) experienced at the end of an experiment. We thus decided to neglect them. As shown in Fig. \ref{Fig2}, the power-law exponent does not change if the pressure loss jump recognition threshold varies (the exponent is calculated for jumps $>$50~\pascal ).

The quiet time distribution of pressure loss jumps exhibits exponential behaviour if the threshold is reduced to 5~\pascal\: (see inset of Fig. \ref{Fig2}).

\begin{figure}[h!]
\includegraphics[width=\linewidth]{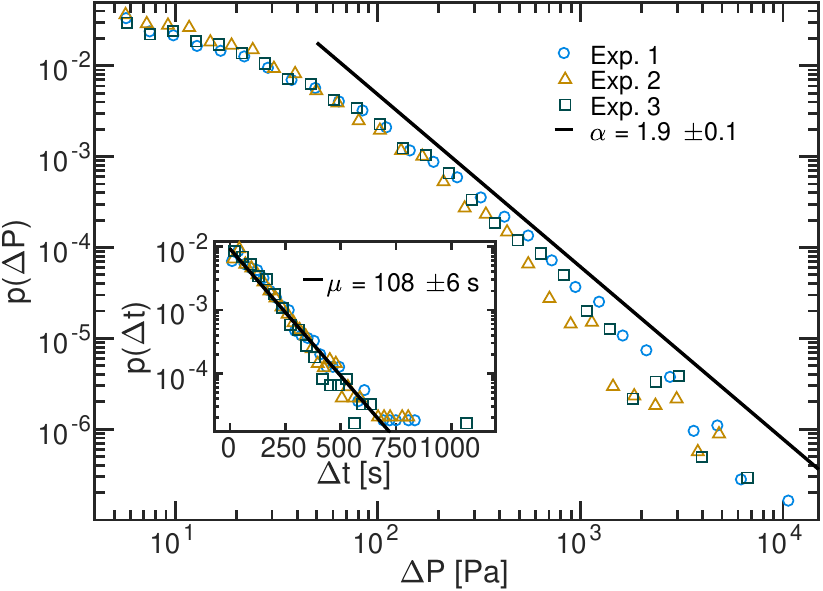}
\caption{Size distribution of pressure loss jumps $\Delta P$ with a recognition threshold of 5~\pascal . It exhibits power-law behaviour with exponent $\alpha$~=~1.9~$\pm$0.1 (black line). The inset shows quiet times distributions. They are fitted with an exponential (black line), whose mean is 108~=~$\pm$6~\second .  The error of the fitted parameters $\alpha$ represents the maximum difference between the fitted parameters of data from all experiments and the fitted parameters of data from a single experiment.}
\label{Fig2}
\end{figure}


\begin{thebibliography}{0}%
\makeatletter
\providecommand \@ifxundefined [1]{%
 \@ifx{#1\undefined}
}%
\providecommand \@ifnum [1]{%
 \ifnum #1\expandafter \@firstoftwo
 \else \expandafter \@secondoftwo
 \fi
}%
\providecommand \@ifx [1]{%
 \ifx #1\expandafter \@firstoftwo
 \else \expandafter \@secondoftwo
 \fi
}%
\providecommand \natexlab [1]{#1}%
\providecommand \enquote  [1]{``#1''}%
\providecommand \bibnamefont  [1]{#1}%
\providecommand \bibfnamefont [1]{#1}%
\providecommand \citenamefont [1]{#1}%
\providecommand \href@noop [0]{\@secondoftwo}%
\providecommand \href [0]{\begingroup \@sanitize@url \@href}%
\providecommand \@href[1]{\@@startlink{#1}\@@href}%
\providecommand \@@href[1]{\endgroup#1\@@endlink}%
\providecommand \@sanitize@url [0]{\catcode `\\12\catcode `\$12\catcode
  `\&12\catcode `\#12\catcode `\^12\catcode `\_12\catcode `\%12\relax}%
\providecommand \@@startlink[1]{}%
\providecommand \@@endlink[0]{}%
\providecommand \url  [0]{\begingroup\@sanitize@url \@url }%
\providecommand \@url [1]{\endgroup\@href {#1}{\urlprefix }}%
\providecommand \urlprefix  [0]{URL }%
\providecommand \Eprint [0]{\href }%
\providecommand \doibase [0]{http://dx.doi.org/}%
\providecommand \selectlanguage [0]{\@gobble}%
\providecommand \bibinfo  [0]{\@secondoftwo}%
\providecommand \bibfield  [0]{\@secondoftwo}%
\providecommand \translation [1]{[#1]}%
\providecommand \BibitemOpen [0]{}%
\providecommand \bibitemStop [0]{}%
\providecommand \bibitemNoStop [0]{.\EOS\space}%
\providecommand \EOS [0]{\spacefactor3000\relax}%
\providecommand \BibitemShut  [1]{\csname bibitem#1\endcsname}%
\let\auto@bib@innerbib\@empty
\end{thebibliography}%


\begin{thebibliography}{43}%
\makeatletter
\providecommand \@ifxundefined [1]{%
 \@ifx{#1\undefined}
}%
\providecommand \@ifnum [1]{%
 \ifnum #1\expandafter \@firstoftwo
 \else \expandafter \@secondoftwo
 \fi
}%
\providecommand \@ifx [1]{%
 \ifx #1\expandafter \@firstoftwo
 \else \expandafter \@secondoftwo
 \fi
}%
\providecommand \natexlab [1]{#1}%
\providecommand \enquote  [1]{``#1''}%
\providecommand \bibnamefont  [1]{#1}%
\providecommand \bibfnamefont [1]{#1}%
\providecommand \citenamefont [1]{#1}%
\providecommand \href@noop [0]{\@secondoftwo}%
\providecommand \href [0]{\begingroup \@sanitize@url \@href}%
\providecommand \@href[1]{\@@startlink{#1}\@@href}%
\providecommand \@@href[1]{\endgroup#1\@@endlink}%
\providecommand \@sanitize@url [0]{\catcode `\\12\catcode `\$12\catcode
  `\&12\catcode `\#12\catcode `\^12\catcode `\_12\catcode `\%12\relax}%
\providecommand \@@startlink[1]{}%
\providecommand \@@endlink[0]{}%
\providecommand \url  [0]{\begingroup\@sanitize@url \@url }%
\providecommand \@url [1]{\endgroup\@href {#1}{\urlprefix }}%
\providecommand \urlprefix  [0]{URL }%
\providecommand \Eprint [0]{\href }%
\providecommand \doibase [0]{http://dx.doi.org/}%
\providecommand \selectlanguage [0]{\@gobble}%
\providecommand \bibinfo  [0]{\@secondoftwo}%
\providecommand \bibfield  [0]{\@secondoftwo}%
\providecommand \translation [1]{[#1]}%
\providecommand \BibitemOpen [0]{}%
\providecommand \bibitemStop [0]{}%
\providecommand \bibitemNoStop [0]{.\EOS\space}%
\providecommand \EOS [0]{\spacefactor3000\relax}%
\providecommand \BibitemShut  [1]{\csname bibitem#1\endcsname}%
\let\auto@bib@innerbib\@empty
\bibitem [{\citenamefont {Gruesbeck}\ and\ \citenamefont
  {Collins}(1982)}]{Gruesbeck1982}%
  \BibitemOpen
  \bibfield  {author} {\bibinfo {author} {\bibfnamefont {C.}~\bibnamefont
  {Gruesbeck}}\ and\ \bibinfo {author} {\bibfnamefont {R.}~\bibnamefont
  {Collins}},\ }\href@noop {} {\bibfield  {journal} {\bibinfo  {journal} {Soc.
  Petrol. Eng. J.}\ } (\bibinfo {year} {1982})}\BibitemShut {NoStop}%
\bibitem [{\citenamefont {Einstein}\ and\ \citenamefont
  {Civan}(1992)}]{Einstein1992}%
  \BibitemOpen
  \bibfield  {author} {\bibinfo {author} {\bibfnamefont {M.~A.}\ \bibnamefont
  {Einstein}}\ and\ \bibinfo {author} {\bibfnamefont {F.}~\bibnamefont
  {Civan}},\ }\href@noop {} {\bibfield  {journal} {\bibinfo  {journal} {J. Can.
  Petrol. Technol.}\ }\textbf {\bibinfo {volume} {31}},\ \bibinfo {pages} {27}
  (\bibinfo {year} {1992})}\BibitemShut {NoStop}%
\bibitem [{\citenamefont {Sahimi}\ \emph {et~al.}(2000)\citenamefont {Sahimi},
  \citenamefont {Mehrabi}, \citenamefont {Mirzaee},\ and\ \citenamefont
  {Rassamdana}}]{Sahimi2000}%
  \BibitemOpen
  \bibfield  {author} {\bibinfo {author} {\bibfnamefont {M.}~\bibnamefont
  {Sahimi}}, \bibinfo {author} {\bibfnamefont {A.~R.}\ \bibnamefont {Mehrabi}},
  \bibinfo {author} {\bibfnamefont {N.}~\bibnamefont {Mirzaee}}, \ and\
  \bibinfo {author} {\bibfnamefont {H.}~\bibnamefont {Rassamdana}},\ }\href
  {\doibase 10.1023/A:1006759524127} {\bibfield  {journal} {\bibinfo  {journal}
  {Transport Porous Med.}\ }\textbf {\bibinfo {volume} {41}},\ \bibinfo {pages}
  {325} (\bibinfo {year} {2000})}\BibitemShut {NoStop}%
\bibitem [{\citenamefont {Herzig}\ \emph {et~al.}(1970)\citenamefont {Herzig},
  \citenamefont {Leclerc},\ and\ \citenamefont {Goff}}]{Herzig1970}%
  \BibitemOpen
  \bibfield  {author} {\bibinfo {author} {\bibfnamefont {J.~P.}\ \bibnamefont
  {Herzig}}, \bibinfo {author} {\bibfnamefont {D.~M.}\ \bibnamefont {Leclerc}},
  \ and\ \bibinfo {author} {\bibfnamefont {P.~L.}\ \bibnamefont {Goff}},\
  }\href {\doibase 10.1021/ie50725a003} {\bibfield  {journal} {\bibinfo
  {journal} {Ind. Eng. Chem.}\ }\textbf {\bibinfo {volume} {62}},\ \bibinfo
  {pages} {8} (\bibinfo {year} {1970})}\BibitemShut {NoStop}%
\bibitem [{\citenamefont {Ken}(2008)}]{Sutherland2008}%
  \BibitemOpen
  \bibfield  {author} {\bibinfo {author} {\bibfnamefont {S.}~\bibnamefont
  {Ken}},\ }\href@noop {} {\emph {\bibinfo {title} {Filters and Filtration
  Handbook}}}\ (\bibinfo  {publisher} {Elsevier},\ \bibinfo {address}
  {Oxford},\ \bibinfo {year} {2008})\BibitemShut {NoStop}%
\bibitem [{\citenamefont {Tien}(2012)}]{Tien2012}%
  \BibitemOpen
  \bibfield  {author} {\bibinfo {author} {\bibfnamefont {C.}~\bibnamefont
  {Tien}},\ }\href@noop {} {\emph {\bibinfo {title} {Principles of
  Filtration}}}\ (\bibinfo  {publisher} {Elsevier},\ \bibinfo {address}
  {Amsterdam},\ \bibinfo {year} {2012})\BibitemShut {NoStop}%
\bibitem [{\citenamefont {Valdes}\ and\ \citenamefont
  {Santamarina}(2006)}]{Valdes2006}%
  \BibitemOpen
  \bibfield  {author} {\bibinfo {author} {\bibfnamefont {J.~R.}\ \bibnamefont
  {Valdes}}\ and\ \bibinfo {author} {\bibfnamefont {J.~C.}\ \bibnamefont
  {Santamarina}},\ }\href@noop {} {\bibfield  {journal} {\bibinfo  {journal}
  {SPE Journal}\ }\textbf {\bibinfo {volume} {11}},\ \bibinfo {pages} {193}
  (\bibinfo {year} {2006})}\BibitemShut {NoStop}%
\bibitem [{\citenamefont {Bai}\ and\ \citenamefont {Tien}(1997)}]{Bai1997}%
  \BibitemOpen
  \bibfield  {author} {\bibinfo {author} {\bibfnamefont {R.}~\bibnamefont
  {Bai}}\ and\ \bibinfo {author} {\bibfnamefont {C.}~\bibnamefont {Tien}},\
  }\href {\doibase http://dx.doi.org/10.1006/jcis.1996.4663} {\bibfield
  {journal} {\bibinfo  {journal} {J. Colloid Interf. Sci.}\ }\textbf {\bibinfo
  {volume} {186}},\ \bibinfo {pages} {307 } (\bibinfo {year}
  {1997})}\BibitemShut {NoStop}%
\bibitem [{\citenamefont {Bergendahl}\ and\ \citenamefont
  {Grasso}(2000)}]{Bergendahl2000}%
  \BibitemOpen
  \bibfield  {author} {\bibinfo {author} {\bibfnamefont {J.}~\bibnamefont
  {Bergendahl}}\ and\ \bibinfo {author} {\bibfnamefont {D.}~\bibnamefont
  {Grasso}},\ }\href {\doibase http://dx.doi.org/10.1016/S0009-2509(99)00422-4}
  {\bibfield  {journal} {\bibinfo  {journal} {Chem. Eng. Sci.}\ }\textbf
  {\bibinfo {volume} {55}},\ \bibinfo {pages} {1523 } (\bibinfo {year}
  {2000})}\BibitemShut {NoStop}%
\bibitem [{\citenamefont {Mahadevan}\ \emph {et~al.}(2012)\citenamefont
  {Mahadevan}, \citenamefont {Orpe}, \citenamefont {Kudrolli},\ and\
  \citenamefont {Mahadevan}}]{Mahadevan2012}%
  \BibitemOpen
  \bibfield  {author} {\bibinfo {author} {\bibfnamefont {A.}~\bibnamefont
  {Mahadevan}}, \bibinfo {author} {\bibfnamefont {A.~V.}\ \bibnamefont {Orpe}},
  \bibinfo {author} {\bibfnamefont {A.}~\bibnamefont {Kudrolli}}, \ and\
  \bibinfo {author} {\bibfnamefont {L.}~\bibnamefont {Mahadevan}},\ }\href
  {http://stacks.iop.org/0295-5075/98/i=5/a=58003} {\bibfield  {journal}
  {\bibinfo  {journal} {EPL (Europhys. Lett.)}\ }\textbf {\bibinfo {volume}
  {98}},\ \bibinfo {pages} {58003} (\bibinfo {year} {2012})}\BibitemShut
  {NoStop}%
\bibitem [{\citenamefont {Han}\ \emph {et~al.}(2009)\citenamefont {Han},
  \citenamefont {Fitzpatrick},\ and\ \citenamefont {Wetherill}}]{Han2009}%
  \BibitemOpen
  \bibfield  {author} {\bibinfo {author} {\bibfnamefont {S.}~\bibnamefont
  {Han}}, \bibinfo {author} {\bibfnamefont {C.~S.}\ \bibnamefont
  {Fitzpatrick}}, \ and\ \bibinfo {author} {\bibfnamefont {A.}~\bibnamefont
  {Wetherill}},\ }\href {\doibase
  http://dx.doi.org/10.1016/j.watres.2008.12.003} {\bibfield  {journal}
  {\bibinfo  {journal} {Water Res.}\ }\textbf {\bibinfo {volume} {43}},\
  \bibinfo {pages} {1171 } (\bibinfo {year} {2009})}\BibitemShut {NoStop}%
\bibitem [{\citenamefont {Kim}\ and\ \citenamefont {Lawler}(2012)}]{Kim2012}%
  \BibitemOpen
  \bibfield  {author} {\bibinfo {author} {\bibfnamefont {J.}~\bibnamefont
  {Kim}}\ and\ \bibinfo {author} {\bibfnamefont {D.~F.}\ \bibnamefont
  {Lawler}},\ }\href {\doibase http://dx.doi.org/10.1016/j.watres.2011.10.059}
  {\bibfield  {journal} {\bibinfo  {journal} {Water Res.}\ }\textbf {\bibinfo
  {volume} {46}},\ \bibinfo {pages} {433 } (\bibinfo {year}
  {2012})}\BibitemShut {NoStop}%
\bibitem [{\citenamefont {Ives}(1989)}]{Ives1989}%
  \BibitemOpen
  \bibfield  {author} {\bibinfo {author} {\bibfnamefont {K.}~\bibnamefont
  {Ives}},\ }\href {\doibase http://dx.doi.org/10.1016/0043-1354(89)90010-9}
  {\bibfield  {journal} {\bibinfo  {journal} {Water Res.}\ }\textbf {\bibinfo
  {volume} {23}},\ \bibinfo {pages} {861 } (\bibinfo {year}
  {1989})}\BibitemShut {NoStop}%
\bibitem [{\citenamefont {Ghidaglia}\ \emph {et~al.}(1996)\citenamefont
  {Ghidaglia}, \citenamefont {de~Arcangelis}, \citenamefont {Hinch},\ and\
  \citenamefont {Guazzelli}}]{Ghidaglia1996}%
  \BibitemOpen
  \bibfield  {author} {\bibinfo {author} {\bibfnamefont {C.}~\bibnamefont
  {Ghidaglia}}, \bibinfo {author} {\bibfnamefont {L.}~\bibnamefont
  {de~Arcangelis}}, \bibinfo {author} {\bibfnamefont {J.}~\bibnamefont
  {Hinch}}, \ and\ \bibinfo {author} {\bibfnamefont {E.}~\bibnamefont
  {Guazzelli}},\ }\href {\doibase 10.1063/1.868810} {\bibfield  {journal}
  {\bibinfo  {journal} {Phys. Fluids}\ }\textbf {\bibinfo {volume} {8}},\
  \bibinfo {pages} {6} (\bibinfo {year} {1996})}\BibitemShut {NoStop}%
\bibitem [{\citenamefont {Kim†}\ \emph {et~al.}(2004)\citenamefont {Kim†},
  ,\ and\ \citenamefont {Tobiason}}]{Kim2004}%
  \BibitemOpen
  \bibfield  {author} {\bibinfo {author} {\bibfnamefont {J.}~\bibnamefont
  {Kim†}}, , \ and\ \bibinfo {author} {\bibfnamefont {J.~E.}\ \bibnamefont
  {Tobiason}},\ }\href {\doibase 10.1021/es0352698} {\bibfield  {journal}
  {\bibinfo  {journal} {Environ. Sci. Technol.}\ }\textbf {\bibinfo {volume}
  {38}},\ \bibinfo {pages} {6132} (\bibinfo {year} {2004})}\BibitemShut
  {NoStop}%
\bibitem [{\citenamefont {Imdakm}\ and\ \citenamefont
  {Sahimi}(1987)}]{Imdakm1987}%
  \BibitemOpen
  \bibfield  {author} {\bibinfo {author} {\bibfnamefont {A.~O.}\ \bibnamefont
  {Imdakm}}\ and\ \bibinfo {author} {\bibfnamefont {M.}~\bibnamefont
  {Sahimi}},\ }\href {\doibase 10.1103/PhysRevA.36.5304} {\bibfield  {journal}
  {\bibinfo  {journal} {Phys. Rev. A}\ }\textbf {\bibinfo {volume} {36}},\
  \bibinfo {pages} {5304} (\bibinfo {year} {1987})}\BibitemShut {NoStop}%
\bibitem [{\citenamefont {Imdakm}\ and\ \citenamefont
  {Sahimi}(1991)}]{Imdakm1991}%
  \BibitemOpen
  \bibfield  {author} {\bibinfo {author} {\bibfnamefont {A.}~\bibnamefont
  {Imdakm}}\ and\ \bibinfo {author} {\bibfnamefont {M.}~\bibnamefont
  {Sahimi}},\ }\href {\doibase http://dx.doi.org/10.1016/0009-2509(91)80158-U}
  {\bibfield  {journal} {\bibinfo  {journal} {Chem. Eng. Sci.}\ }\textbf
  {\bibinfo {volume} {46}},\ \bibinfo {pages} {1977 } (\bibinfo {year}
  {1991})}\BibitemShut {NoStop}%
\bibitem [{\citenamefont {Sahimi}\ and\ \citenamefont
  {Imdakm}(1991)}]{Sahimi1991}%
  \BibitemOpen
  \bibfield  {author} {\bibinfo {author} {\bibfnamefont {M.}~\bibnamefont
  {Sahimi}}\ and\ \bibinfo {author} {\bibfnamefont {A.~O.}\ \bibnamefont
  {Imdakm}},\ }\href {\doibase 10.1103/PhysRevLett.66.1169} {\bibfield
  {journal} {\bibinfo  {journal} {Phys. Rev. Lett.}\ }\textbf {\bibinfo
  {volume} {66}},\ \bibinfo {pages} {1169} (\bibinfo {year}
  {1991})}\BibitemShut {NoStop}%
\bibitem [{\citenamefont {Ohen}\ and\ \citenamefont {Civan}(1993)}]{Ohen1993}%
  \BibitemOpen
  \bibfield  {author} {\bibinfo {author} {\bibfnamefont {H.~A.}\ \bibnamefont
  {Ohen}}\ and\ \bibinfo {author} {\bibfnamefont {F.}~\bibnamefont {Civan}},\
  }\href@noop {} {\bibfield  {journal} {\bibinfo  {journal} {SPE Advanced
  Technology Series}\ }\textbf {\bibinfo {volume} {1}},\ \bibinfo {pages} {27}
  (\bibinfo {year} {1993})}\BibitemShut {NoStop}%
\bibitem [{\citenamefont {Ju}\ \emph {et~al.}(2007)\citenamefont {Ju},
  \citenamefont {Fan}, \citenamefont {Wang},\ and\ \citenamefont
  {Qiu}}]{Ju2007}%
  \BibitemOpen
  \bibfield  {author} {\bibinfo {author} {\bibfnamefont {B.}~\bibnamefont
  {Ju}}, \bibinfo {author} {\bibfnamefont {T.}~\bibnamefont {Fan}}, \bibinfo
  {author} {\bibfnamefont {X.}~\bibnamefont {Wang}}, \ and\ \bibinfo {author}
  {\bibfnamefont {X.}~\bibnamefont {Qiu}},\ }\href {\doibase
  10.1007/s11242-006-9044-9} {\bibfield  {journal} {\bibinfo  {journal}
  {Transport Porous Med.}\ }\textbf {\bibinfo {volume} {68}},\ \bibinfo {pages}
  {265} (\bibinfo {year} {2007})}\BibitemShut {NoStop}%
\bibitem [{\citenamefont {Civan}(2007)}]{Civan2007}%
  \BibitemOpen
  \bibfield  {author} {\bibinfo {author} {\bibfnamefont {F.}~\bibnamefont
  {Civan}},\ }\href@noop {} {\emph {\bibinfo {title} {Reservoir formation
  damage: fundamentals, modeling, assessment, and mitigation}}}\ (\bibinfo
  {publisher} {Gulf Professional Publishing},\ \bibinfo {year}
  {2007})\BibitemShut {NoStop}%
\bibitem [{\citenamefont {Ochi}\ and\ \citenamefont
  {Vernoux}(1999)}]{Ochi1999}%
  \BibitemOpen
  \bibfield  {author} {\bibinfo {author} {\bibfnamefont {J.}~\bibnamefont
  {Ochi}}\ and\ \bibinfo {author} {\bibfnamefont {J.-F.}\ \bibnamefont
  {Vernoux}},\ }\href {\doibase 10.1023/A:1006690700000} {\bibfield  {journal}
  {\bibinfo  {journal} {Transport Porous Med.}\ }\textbf {\bibinfo {volume}
  {37}},\ \bibinfo {pages} {303} (\bibinfo {year} {1999})}\BibitemShut
  {NoStop}%
\bibitem [{\citenamefont {Sahimi}\ \emph {et~al.}(1990)\citenamefont {Sahimi},
  \citenamefont {Gavalas},\ and\ \citenamefont {Tsotsis}}]{Sahimi1990}%
  \BibitemOpen
  \bibfield  {author} {\bibinfo {author} {\bibfnamefont {M.}~\bibnamefont
  {Sahimi}}, \bibinfo {author} {\bibfnamefont {G.~R.}\ \bibnamefont {Gavalas}},
  \ and\ \bibinfo {author} {\bibfnamefont {T.~T.}\ \bibnamefont {Tsotsis}},\
  }\href {\doibase http://dx.doi.org/10.1016/0009-2509(90)80001-U} {\bibfield
  {journal} {\bibinfo  {journal} {Chem. Eng. Sci.}\ }\textbf {\bibinfo {volume}
  {45}},\ \bibinfo {pages} {1443 } (\bibinfo {year} {1990})}\BibitemShut
  {NoStop}%
\bibitem [{\citenamefont {Newman}(2005)}]{Newman2005}%
  \BibitemOpen
  \bibfield  {author} {\bibinfo {author} {\bibfnamefont {M.~E.~J.}\
  \bibnamefont {Newman}},\ }\href {\doibase 10.1080/00107510500052444}
  {\bibfield  {journal} {\bibinfo  {journal} {Contemp. Phys.}\ }\textbf
  {\bibinfo {volume} {46}},\ \bibinfo {pages} {323} (\bibinfo {year}
  {2005})}\BibitemShut {NoStop}%
\bibitem [{\citenamefont {J\"ager}\ \emph {et~al.}()\citenamefont {J\"ager},
  \citenamefont {Mendoza},\ and\ \citenamefont {Herrmann}}]{Jaeger2017}%
  \BibitemOpen
  \bibfield  {author} {\bibinfo {author} {\bibfnamefont {R.}~\bibnamefont
  {J\"ager}}, \bibinfo {author} {\bibfnamefont {M.}~\bibnamefont {Mendoza}}, \
  and\ \bibinfo {author} {\bibfnamefont {H.~J.}\ \bibnamefont {Herrmann}},\
  }\href@noop {} {\enquote {\bibinfo {title} {The mechanism behind erosive
  bursts in porous media},}\ }\bibinfo {note} {To appear on Phys. Rev.
  Lett.}\BibitemShut {Stop}%
\bibitem [{\citenamefont {de~Arcangelis}\ \emph {et~al.}(2016)\citenamefont
  {de~Arcangelis}, \citenamefont {Godano}, \citenamefont {Grasso},\ and\
  \citenamefont {Lippiello}}]{DeArcangelis2016}%
  \BibitemOpen
  \bibfield  {author} {\bibinfo {author} {\bibfnamefont {L.}~\bibnamefont
  {de~Arcangelis}}, \bibinfo {author} {\bibfnamefont {C.}~\bibnamefont
  {Godano}}, \bibinfo {author} {\bibfnamefont {J.~R.}\ \bibnamefont {Grasso}},
  \ and\ \bibinfo {author} {\bibfnamefont {E.}~\bibnamefont {Lippiello}},\
  }\href {\doibase http://dx.doi.org/10.1016/j.physrep.2016.03.002} {\bibfield
  {journal} {\bibinfo  {journal} {Phys. Rep.}\ }\textbf {\bibinfo {volume}
  {628}},\ \bibinfo {pages} {1 } (\bibinfo {year} {2016})}\BibitemShut
  {NoStop}%
\bibitem [{\citenamefont {Mendoza}\ \emph {et~al.}(2014)\citenamefont
  {Mendoza}, \citenamefont {Kaydul}, \citenamefont {de~Arcangelis},
  \citenamefont {Andrade~Jr},\ and\ \citenamefont {Herrmann}}]{Mendoza2014}%
  \BibitemOpen
  \bibfield  {author} {\bibinfo {author} {\bibfnamefont {M.}~\bibnamefont
  {Mendoza}}, \bibinfo {author} {\bibfnamefont {A.}~\bibnamefont {Kaydul}},
  \bibinfo {author} {\bibfnamefont {L.}~\bibnamefont {de~Arcangelis}}, \bibinfo
  {author} {\bibfnamefont {J.~S.}\ \bibnamefont {Andrade~Jr}}, \ and\ \bibinfo
  {author} {\bibfnamefont {H.~J.}\ \bibnamefont {Herrmann}},\ }\href@noop {}
  {\bibfield  {journal} {\bibinfo  {journal} {Nat. Commun.}\ }\textbf {\bibinfo
  {volume} {5}},\ \bibinfo {pages} {5035} (\bibinfo {year} {2014})}\BibitemShut
  {NoStop}%
\bibitem [{\citenamefont {Beggs}\ and\ \citenamefont
  {Plenz}(2003)}]{Beggs2003}%
  \BibitemOpen
  \bibfield  {author} {\bibinfo {author} {\bibfnamefont {J.~M.}\ \bibnamefont
  {Beggs}}\ and\ \bibinfo {author} {\bibfnamefont {D.}~\bibnamefont {Plenz}},\
  }\href@noop {} {\bibfield  {journal} {\bibinfo  {journal} {J. Neurosci.}\
  }\textbf {\bibinfo {volume} {23}},\ \bibinfo {pages} {11167} (\bibinfo {year}
  {2003})}\BibitemShut {NoStop}%
\bibitem [{\citenamefont {Corral}(2003)}]{Corral2003}%
  \BibitemOpen
  \bibfield  {author} {\bibinfo {author} {\bibfnamefont {A.}~\bibnamefont
  {Corral}},\ }\href {\doibase 10.1103/PhysRevE.68.035102} {\bibfield
  {journal} {\bibinfo  {journal} {Phys. Rev. E}\ }\textbf {\bibinfo {volume}
  {68}},\ \bibinfo {pages} {035102} (\bibinfo {year} {2003})}\BibitemShut
  {NoStop}%
\bibitem [{\citenamefont {Corral}(2004)}]{Corral2004}%
  \BibitemOpen
  \bibfield  {author} {\bibinfo {author} {\bibfnamefont {A.}~\bibnamefont
  {Corral}},\ }\href {\doibase 10.1103/PhysRevLett.92.108501} {\bibfield
  {journal} {\bibinfo  {journal} {Phys. Rev. Lett.}\ }\textbf {\bibinfo
  {volume} {92}},\ \bibinfo {pages} {108501} (\bibinfo {year}
  {2004})}\BibitemShut {NoStop}%
\bibitem [{\citenamefont {Crosby}\ \emph {et~al.}(1998)\citenamefont {Crosby},
  \citenamefont {Vilmer}, \citenamefont {Lund},\ and\ \citenamefont
  {Sunyaev}}]{Crosby1998}%
  \BibitemOpen
  \bibfield  {author} {\bibinfo {author} {\bibfnamefont {N.}~\bibnamefont
  {Crosby}}, \bibinfo {author} {\bibfnamefont {N.}~\bibnamefont {Vilmer}},
  \bibinfo {author} {\bibfnamefont {N.}~\bibnamefont {Lund}}, \ and\ \bibinfo
  {author} {\bibfnamefont {R.}~\bibnamefont {Sunyaev}},\ }\href@noop {}
  {\bibfield  {journal} {\bibinfo  {journal} {Astron. Astrophys.}\ }\textbf
  {\bibinfo {volume} {334}},\ \bibinfo {pages} {299} (\bibinfo {year}
  {1998})}\BibitemShut {NoStop}%
\bibitem [{\citenamefont {Wheatland}\ \emph {et~al.}(1998)\citenamefont
  {Wheatland}, \citenamefont {Sturrock},\ and\ \citenamefont
  {McTiernan}}]{Wheatland1998}%
  \BibitemOpen
  \bibfield  {author} {\bibinfo {author} {\bibfnamefont {M.~S.}\ \bibnamefont
  {Wheatland}}, \bibinfo {author} {\bibfnamefont {P.~A.}\ \bibnamefont
  {Sturrock}}, \ and\ \bibinfo {author} {\bibfnamefont {J.~M.}\ \bibnamefont
  {McTiernan}},\ }\href@noop {} {\bibfield  {journal} {\bibinfo  {journal}
  {Astrophys. J.}\ }\textbf {\bibinfo {volume} {509}},\ \bibinfo {pages} {448}
  (\bibinfo {year} {1998})}\BibitemShut {NoStop}%
\bibitem [{\citenamefont {Davidsen}\ \emph {et~al.}(2007)\citenamefont
  {Davidsen}, \citenamefont {Stanchits},\ and\ \citenamefont
  {Dresen}}]{Davidsen2007}%
  \BibitemOpen
  \bibfield  {author} {\bibinfo {author} {\bibfnamefont {J.}~\bibnamefont
  {Davidsen}}, \bibinfo {author} {\bibfnamefont {S.}~\bibnamefont {Stanchits}},
  \ and\ \bibinfo {author} {\bibfnamefont {G.}~\bibnamefont {Dresen}},\ }\href
  {\doibase 10.1103/PhysRevLett.98.125502} {\bibfield  {journal} {\bibinfo
  {journal} {Phys. Rev. Lett.}\ }\textbf {\bibinfo {volume} {98}},\ \bibinfo
  {pages} {125502} (\bibinfo {year} {2007})}\BibitemShut {NoStop}%
\bibitem [{\citenamefont {de~Arcangelis}\ \emph {et~al.}(2006)\citenamefont
  {de~Arcangelis}, \citenamefont {Godano}, \citenamefont {Lippiello},\ and\
  \citenamefont {Nicodemi}}]{DeArcangelis2006}%
  \BibitemOpen
  \bibfield  {author} {\bibinfo {author} {\bibfnamefont {L.}~\bibnamefont
  {de~Arcangelis}}, \bibinfo {author} {\bibfnamefont {C.}~\bibnamefont
  {Godano}}, \bibinfo {author} {\bibfnamefont {E.}~\bibnamefont {Lippiello}}, \
  and\ \bibinfo {author} {\bibfnamefont {M.}~\bibnamefont {Nicodemi}},\ }\href
  {\doibase 10.1103/PhysRevLett.96.051102} {\bibfield  {journal} {\bibinfo
  {journal} {Phys. Rev. Lett.}\ }\textbf {\bibinfo {volume} {96}},\ \bibinfo
  {pages} {051102} (\bibinfo {year} {2006})}\BibitemShut {NoStop}%
\bibitem [{\citenamefont {Utsu}\ \emph {et~al.}(1995)\citenamefont {Utsu},
  \citenamefont {Ogata}, \citenamefont {S},\ and\ \citenamefont
  {Matsu'ura}}]{Utsu1995}%
  \BibitemOpen
  \bibfield  {author} {\bibinfo {author} {\bibfnamefont {T.}~\bibnamefont
  {Utsu}}, \bibinfo {author} {\bibfnamefont {Y.}~\bibnamefont {Ogata}},
  \bibinfo {author} {\bibfnamefont {R.}~\bibnamefont {S}}, \ and\ \bibinfo
  {author} {\bibnamefont {Matsu'ura}},\ }\href {\doibase 10.4294/jpe1952.43.1}
  {\bibfield  {journal} {\bibinfo  {journal} {J. Phys. Earth}\ }\textbf
  {\bibinfo {volume} {43}},\ \bibinfo {pages} {1} (\bibinfo {year}
  {1995})}\BibitemShut {NoStop}%
\bibitem [{\citenamefont {Main}(2000)}]{Main2000}%
  \BibitemOpen
  \bibfield  {author} {\bibinfo {author} {\bibfnamefont {I.~G.}\ \bibnamefont
  {Main}},\ }\href {\doibase 10.1046/j.1365-246x.2000.00136.x} {\bibfield
  {journal} {\bibinfo  {journal} {Geophys. J. Int.}\ }\textbf {\bibinfo
  {volume} {142}},\ \bibinfo {pages} {151} (\bibinfo {year}
  {2000})}\BibitemShut {NoStop}%
\bibitem [{\citenamefont {Ogata}(1983)}]{Ogata1983}%
  \BibitemOpen
  \bibfield  {author} {\bibinfo {author} {\bibfnamefont {Y.}~\bibnamefont
  {Ogata}},\ }\href {\doibase 10.4294/jpe1952.31.115} {\bibfield  {journal}
  {\bibinfo  {journal} {J. Phys. Earth}\ }\textbf {\bibinfo {volume} {31}},\
  \bibinfo {pages} {115} (\bibinfo {year} {1983})}\BibitemShut {NoStop}%
\bibitem [{\citenamefont {Omori}(1894)}]{Omori1894}%
  \BibitemOpen
  \bibfield  {author} {\bibinfo {author} {\bibfnamefont {F.}~\bibnamefont
  {Omori}},\ }\href@noop {} {\bibfield  {journal} {\bibinfo  {journal} {J.
  Coll. Sci. Imp. Univ. Tokyo}\ }\textbf {\bibinfo {volume} {7}},\ \bibinfo
  {pages} {111} (\bibinfo {year} {1894})}\BibitemShut {NoStop}%
\bibitem [{\citenamefont {Gutenberg}\ and\ \citenamefont
  {Richter}(1944)}]{Gutenberg1944}%
  \BibitemOpen
  \bibfield  {author} {\bibinfo {author} {\bibfnamefont {B.}~\bibnamefont
  {Gutenberg}}\ and\ \bibinfo {author} {\bibfnamefont {C.~F.}\ \bibnamefont
  {Richter}},\ }\href {http://www.bssaonline.org/content/34/4/185.short}
  {\bibfield  {journal} {\bibinfo  {journal} {B. Seismol. Soc. Am.}\ }\textbf
  {\bibinfo {volume} {34}},\ \bibinfo {pages} {185} (\bibinfo {year}
  {1944})}\BibitemShut {NoStop}%
\bibitem [{\citenamefont {Lu}\ and\ \citenamefont {Hamilton}(1991)}]{Lu1991}%
  \BibitemOpen
  \bibfield  {author} {\bibinfo {author} {\bibfnamefont {E.~T.}\ \bibnamefont
  {Lu}}\ and\ \bibinfo {author} {\bibfnamefont {R.~J.}\ \bibnamefont
  {Hamilton}},\ }\href@noop {} {\bibfield  {journal} {\bibinfo  {journal}
  {Astrophys. J.}\ }\textbf {\bibinfo {volume} {380}},\ \bibinfo {pages} {L89}
  (\bibinfo {year} {1991})}\BibitemShut {NoStop}%
\bibitem [{\citenamefont {Gopikrishnan}\ \emph {et~al.}(2000)\citenamefont
  {Gopikrishnan}, \citenamefont {Plerou}, \citenamefont {Gabaix},\ and\
  \citenamefont {Stanley}}]{Gopikrishnan2000}%
  \BibitemOpen
  \bibfield  {author} {\bibinfo {author} {\bibfnamefont {P.}~\bibnamefont
  {Gopikrishnan}}, \bibinfo {author} {\bibfnamefont {V.}~\bibnamefont
  {Plerou}}, \bibinfo {author} {\bibfnamefont {X.}~\bibnamefont {Gabaix}}, \
  and\ \bibinfo {author} {\bibfnamefont {H.~E.}\ \bibnamefont {Stanley}},\
  }\href {\doibase 10.1103/PhysRevE.62.R4493} {\bibfield  {journal} {\bibinfo
  {journal} {Phys. Rev. E}\ }\textbf {\bibinfo {volume} {62}},\ \bibinfo
  {pages} {R4493} (\bibinfo {year} {2000})}\BibitemShut {NoStop}%
\bibitem [{\citenamefont {Bartolozzi}\ \emph {et~al.}(2005)\citenamefont
  {Bartolozzi}, \citenamefont {Leinweber},\ and\ \citenamefont
  {Thomas}}]{Bartolozzi2005}%
  \BibitemOpen
  \bibfield  {author} {\bibinfo {author} {\bibfnamefont {M.}~\bibnamefont
  {Bartolozzi}}, \bibinfo {author} {\bibfnamefont {D.}~\bibnamefont
  {Leinweber}}, \ and\ \bibinfo {author} {\bibfnamefont {A.}~\bibnamefont
  {Thomas}},\ }\href {\doibase http://dx.doi.org/10.1016/j.physa.2004.11.061}
  {\bibfield  {journal} {\bibinfo  {journal} {Physica A}\ }\textbf {\bibinfo
  {volume} {350}},\ \bibinfo {pages} {451 } (\bibinfo {year}
  {2005})}\BibitemShut {NoStop}%
\bibitem [{\citenamefont {de~Arcangelis}\ and\ \citenamefont
  {Herrmann}(2014)}]{DeArcangelis2014}%
  \BibitemOpen
  \bibfield  {author} {\bibinfo {author} {\bibfnamefont {L.}~\bibnamefont
  {de~Arcangelis}}\ and\ \bibinfo {author} {\bibfnamefont {H.~J.}\ \bibnamefont
  {Herrmann}},\ }in\ \href {\doibase 10.1002/9783527651009.ch12} {\emph
  {\bibinfo {booktitle} {Criticality in Neural Systems}}},\ \bibinfo {editor}
  {edited by\ \bibinfo {editor} {\bibfnamefont {D.}~\bibnamefont {Plenz}}\ and\
  \bibinfo {editor} {\bibfnamefont {N.}~\bibnamefont {Ernst}}}\ (\bibinfo
  {publisher} {Wiley-VCH Verlag GmbH \& Co. KGaA},\ \bibinfo {address}
  {Weinheim},\ \bibinfo {year} {2014})\ pp.\ \bibinfo {pages}
  {273--292}\BibitemShut {NoStop}%
\end{thebibliography}
\end{document}